\documentclass[12pt,thmsa,a4paper]{article}
\usepackage{sw20lart}
\usepackage{amsmath}
\usepackage{graphicx}
\usepackage{amsfonts}
\usepackage{amssymb}

\begin{document}

\author{S. Beirle, U. Werthenbach, G. Zech, T. Zeuner\\Universit\"{a}t Siegen, Germany}
\title{Carbon Coated Gas Electron Multipliers}
\date{July 1998}
\maketitle
\begin{abstract}
Gas electron multipliers (GEMs) have been overcoated with a high resistivity
10$^{14}-10^{15}\Omega/\square$ amorphous carbon layer. The coating avoids
charging up of the holes and provides a constant gain immediately after
switching on independent of the rate. The gain uniformity across the GEM is
improved. Coating opens the possibility to produce thick GEMs of very high gain.
\end{abstract}

\section{Introduction}

The gas electron multiplier (GEM) [1,2] has proven to be a very attractive new
device especially in combination with MSGCs. Splitting the amplification in
two steps allows to reduce the MSGC voltage to values safely below sparking
thresholds. The functionality of GEM-MSGCs has been demonstrated in hadronic
beams [3] and in magnetic fields up to 1 Tesla. Gains in excess of 10$^{3}$
have been reached with GEMs [4] which open the perspective to replace the MSGC
by a simple passive anode structure. Many interesting results have been
published by the CERN group [4,5] and the large scale application of the GEM
is foreseen in the HERA-B experiment.[3,6]

Like in all gaseous detectors, the maximum gain of GEMs is limited by
discharges. Breakdown may occur through streamer discharge due to large space
charge densities. Gains of about 10$^{3}$ can be achieved in one amplification
step depending on the strength and shape of the electric field [7]. In the
case of GEMs or similar devices a large amplification gap with low gain per
unit length is preferable to a high condensed field. Another limiting factor
is a possible non-uniformity of the field caused by sharp electrode edges or
variable resistivity of insulating surfaces. The latter certainly limit the
maximum distance of the two conducting planes. Indirect indications for
surface discharges in GEMs have been reported [8]. The inhomogeneities in the
surface resistivities can be avoided by coating the surface with a slightly
conducting layer, a procedure which has been very successful with MSGCs.
Furthermore such a coating eliminates gain variations due to polarization of
the plastic and to charging up by ion deposition. These unpleasant effects
have been observed in MSGCs on plastic substrates [9] and led to abandon these
devices. To avoid charging-up it was proposed to add a small amount of water
to the gas [2] which, however, may lead to fast aging [3]. The CERN group
reported measurements using GEMs with cylindrical holes where gain variations
were totally absent [4].

The following measurements have been performed with GEMs which were overcoated
with an amorphous carbon layer. The behavior before and after the coating was compared.

\section{The setup}

The GEMs have been produced at the CERN workshop and coated by a standard
industrial CVD process\footnote{a-Si:C:H:N coating : Fraunhofer Gesellschaft
f\"{u}r Schicht und Oberfl\"{a}chentechnik, Braunschweig, Germany}. The active
GEM area was $35\times27mm^{2}$ (type M) and $100\times100mm^{2}$ (type L).
The $50\mu m$ thick polyimide foil was covered on both sides by $15\mu m$
copper, some of the foils have only $4\mu m$ copper because of a second
etching process. GEM foils of type M have a square hole lattice with a pitch
of $140\mu m$, GEMs of type L have a staggered structure with a minimum
distance between two holes (center to center) of $140\mu m$. The hole
diameters varied between $45$ and $90\mu m$ at the narrowest part of the
Kapton and $70$ to $110\mu m$ at the copper layers. A top view of typical GEM
foils is shown in figure 1.%
\begin{figure}
[ptb]
\begin{center}
\includegraphics[
height=18.1419cm,
width=12.5867cm
]%
{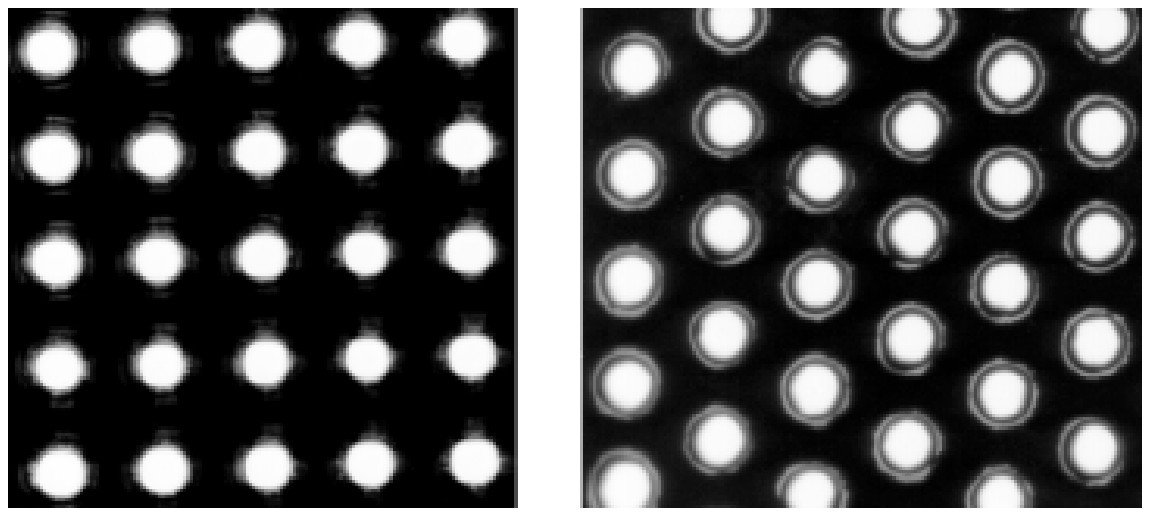}%
\caption{Top view of GEM foils. Type M with rectangular shaped holes and
square hole lattice (left) and type L with round holes and staggered hole
structure (right).}%
\end{center}
\end{figure}

The quality of the first small series (type M) was exceptionally poor with
large variations of the hole diameters of about $30\%$ and many defects. All
the other GEM foils we investigated were of high quality like those shown in
the figure.

The GEMs were glued onto G10 frames and inserted in MSGCs constructed in such
a way as to allow an easy exchange of the GEMs. With the gluing of the GEMs
onto frames special care has to be taken to the flatness of the GEMs.
Variations in the distance between MSGC structure and GEM foil lead to large
gain variations.

The MSGC structure had $300\mu m$ pitch, the anodes and cathodes were $10$ and
$170\mu m$ wide. The gas volume of $6mm$ height was subdivided into two equal
halves by the GEM. We operated the MSGC+GEM with a gas mixture of Ar/DME of
50:50, a variable GEM voltage below $700V$ and a drift field of about
$3.5kV/cm$. The MSGC anodes were grounded. The voltages were chosen such that
the drift field below the GEM foil was slightly higher than above. The chamber
was irradiated with Fe-55 and an $\alpha$ source (Ra-226).

The active area of the MSGC+GEM was scanned with a computer controlled
moveable table and a collimated Fe-55 source.

\section{Results}

\subsection{Electrical properties of the GEMs}

We have measured the electric resistance of the GEMs before and after the
coating. The electrical resistance of the small GEMs was about $10^{13}\Omega$
before and $10^{10}\Omega$ after the coating. The apparent resistivity of the
uncoated polyimide increases by about one and a half orders of magnitude
during the polarization process of about one day duration. Figure 2 compares
its variation with time to the coated version which is completely stable.
Notice the different scales.%
\begin{figure}
[ptb]
\begin{center}
\includegraphics[
height=18.1419cm,
width=12.5867cm
]%
{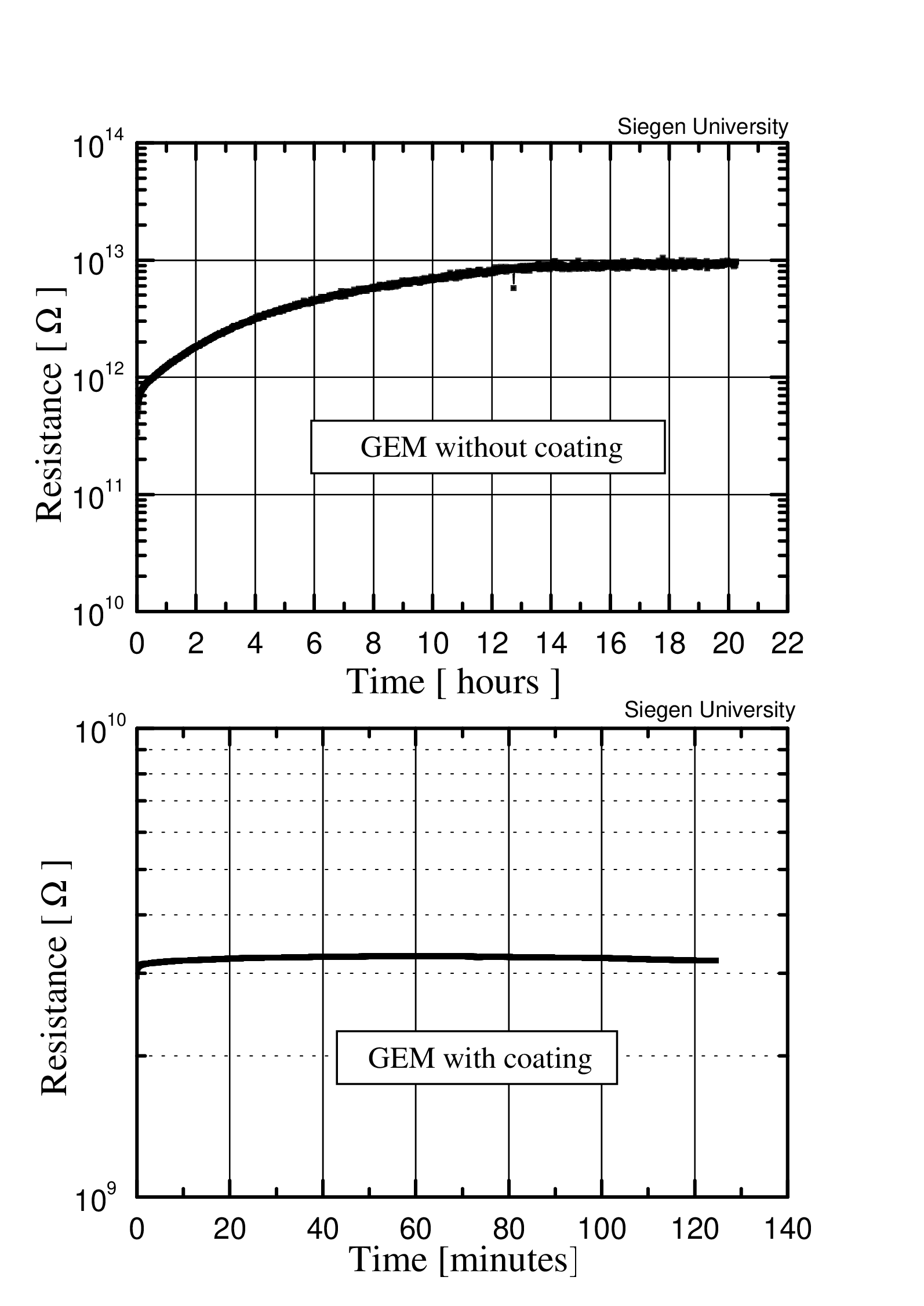}%
\caption{Electrical resistance of GEMs without and with coating (type M).}%
\label{widerstand}%
\end{center}
\end{figure}
For the larger GEMs (type L) the resistance before coating was about
$10^{14}\Omega$. The reason for the higher resistance compared to the small
GEMs is not fully understood. A reason may be a different polyimide material
or different processing. After the coating the resistance of the large GEMs
was of the order of $10^{12}\Omega.$ (The resistivity of the coating was
increased to reduce dark currents.)

\subsection{Energy resolution}

As mentioned before, the GEMs of the first small series (type M) show large
variations of the hole diameters. For this reason the energy resolution of the
uncoated GEMs of this type is rather poor. This is illustrated in Figure 3
where the iron spectrum is compared to that of the coated GEM. In both cases
the irradiated area is about $20mm^{2}$. Scanning with a strongly collimated
source we measured gain variation by a factor two whereas they are in the
range of about 10\% for the coated GEM (Figure 4).%
\begin{figure}
[ptb]
\begin{center}
\includegraphics[
height=16.9909cm,
width=11.7893cm
]%
{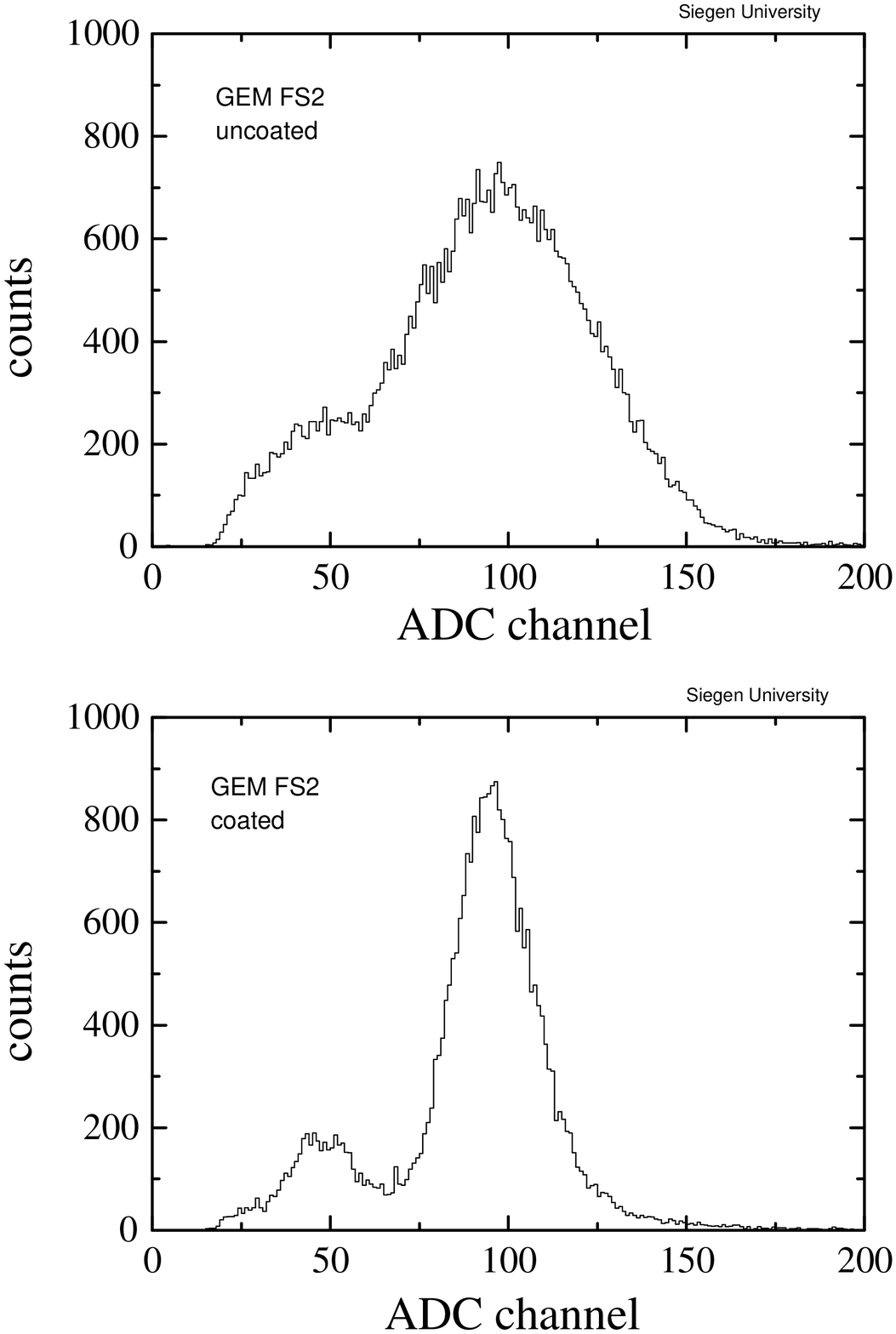}%
\caption{Pulse height spectra from a Fe-55 source for an uncoated and a coated
MSGC+GEM.}%
\label{vgl_spek}%
\end{center}
\end{figure}
\begin{figure}
[ptbptb]
\begin{center}
\includegraphics[
height=18.1419cm,
width=12.5867cm
]%
{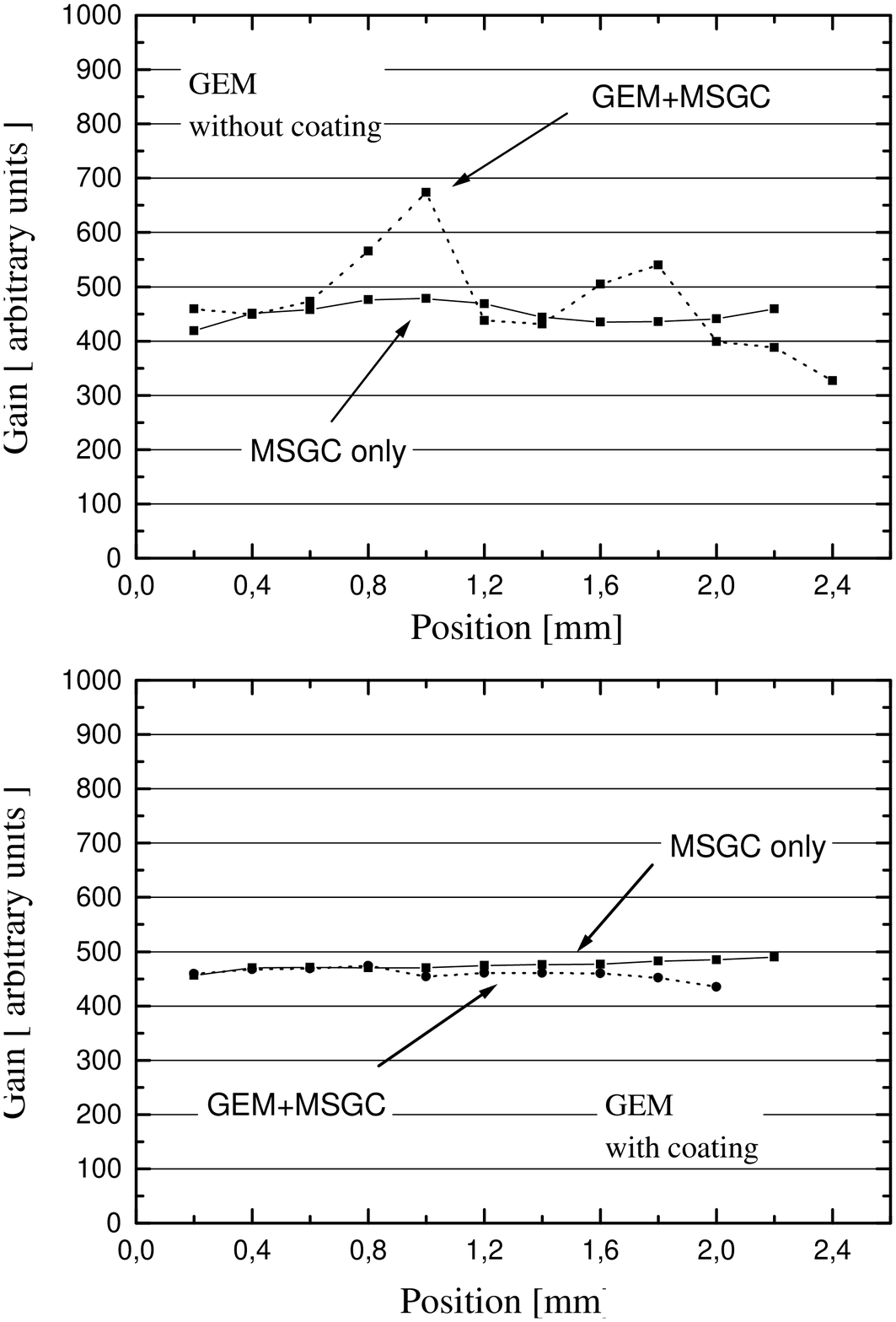}%
\caption{Spatial gain uniformity for an uncoated and a coated MSGC+GEM.}%
\label{vgl_ort}%
\end{center}
\end{figure}
The superior behavior of the coated GEM is probably explained by its more
homogenous field.

The large high quality GEMs (type L) showed comparable energy resolutions and
gain variations before and after coating.

\subsection{Gain stability}

For the uncoated GEMs drastic gain variations due to polarization of the
Kapton and to charging up by ion deposition are observed. Figure 5%
\begin{figure}
[ptb]
\begin{center}
\includegraphics[
height=18.1463cm,
width=12.9733cm
]%
{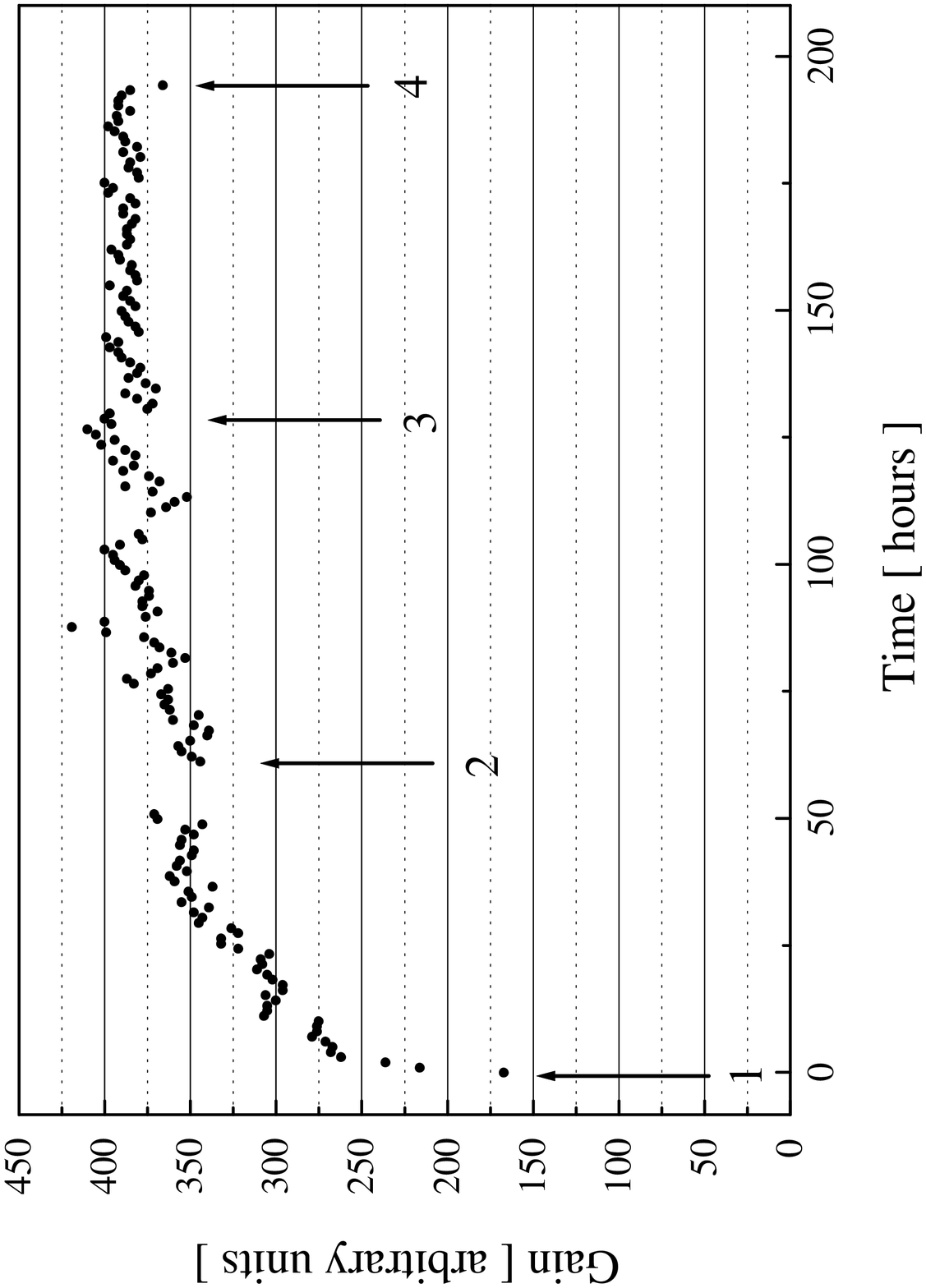}%
\caption{Gain of a MSGC+GEM as a function of time for an uncoated GEM.}%
\label{uncoated_lang}%
\end{center}
\end{figure}
shows the gain as a function of time. The gain increases by about a factor of
two within 40 hours after switching on the HV. To disentangle the effects of
polarization and charging-up we have measured the gain across the active GEM
area after different time intervals. The GEM was irradiated at a fixed
position near its center. The results are displayed in Figure 6.%
\begin{figure}
[ptbptb]
\begin{center}
\includegraphics[
height=18.0738cm,
width=12.8041cm
]%
{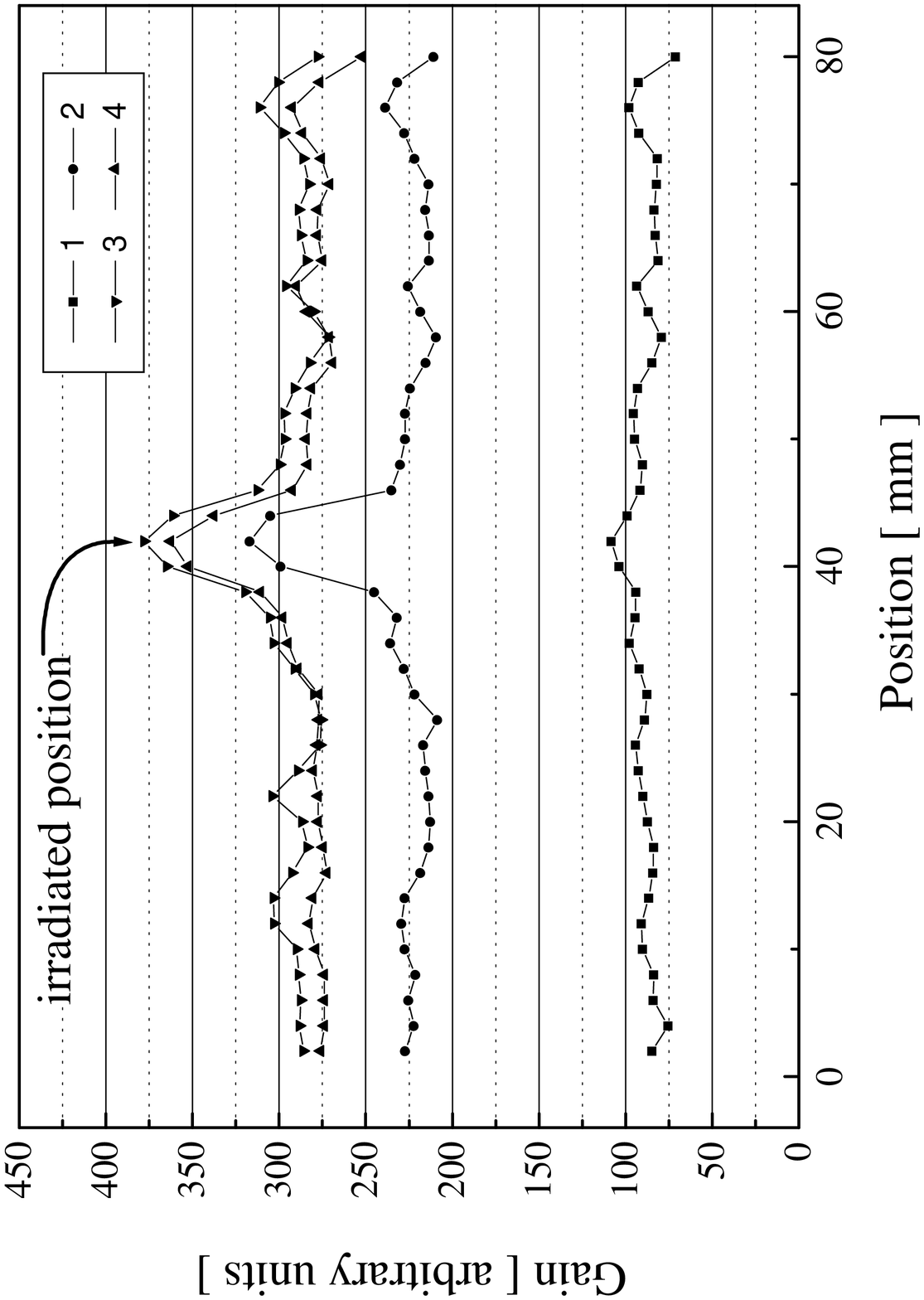}%
\caption{Spatial gain uniformity of a coated GEM at different time intervals
(see Fig. 5) after switching on HV. The gain is higher at the irradiated spot
and remains high for several days after the source has been removed.}%
\label{scan_uncoated}%
\end{center}
\end{figure}
Initially the gain was rather uniform across the GEM (curve 1). After
$\approx60$ hours (curve 2) the gain has globally increased but at the
irradiated spot an additional 40\% increase is observed. The uniform gain
increase of the whole GEM is explained by polarization of the polyimide
whereas the additional local increase at the irradiated spot is due to
charging-up by ion deposition. The gain remains high for several days after
the source has been removed.

The gain of coated GEMs is nearly stable directly after switching on the HV.
This is illustrated in Figure 7.%
\begin{figure}
[ptb]
\begin{center}
\includegraphics[
height=18.1397cm,
width=14.0804cm
]%
{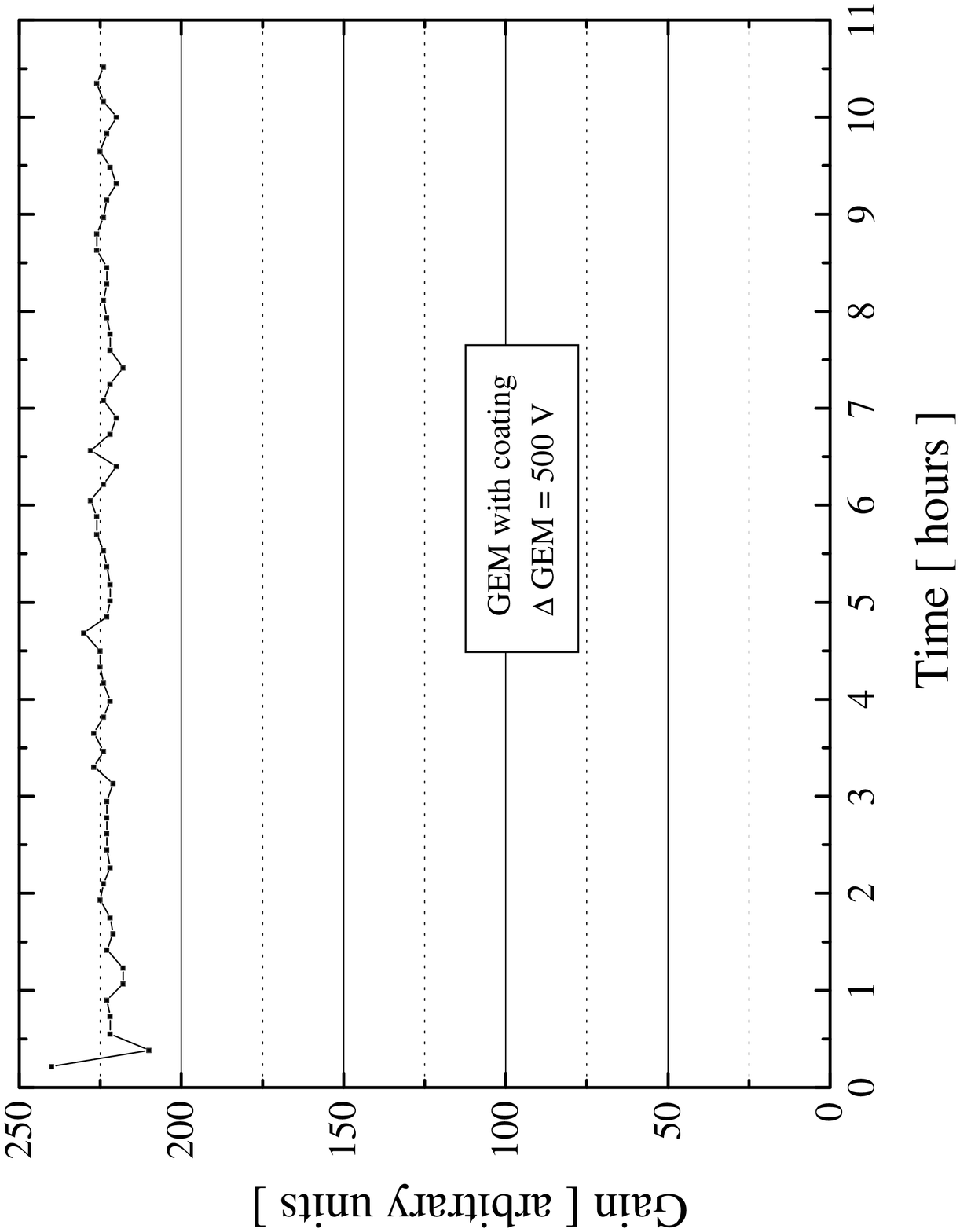}%
\caption{Gain of a coated MSGC+GEM as a function of time.}%
\label{einschalt_coated}%
\end{center}
\end{figure}

\subsection{Maximum achievable Gain}

Figure 8 shows the gas amplification as a function of the cathode voltage for
different constant GEM voltages.%
\begin{figure}
[ptb]
\begin{center}
\includegraphics[
height=17.9574cm,
width=12.9733cm
]%
{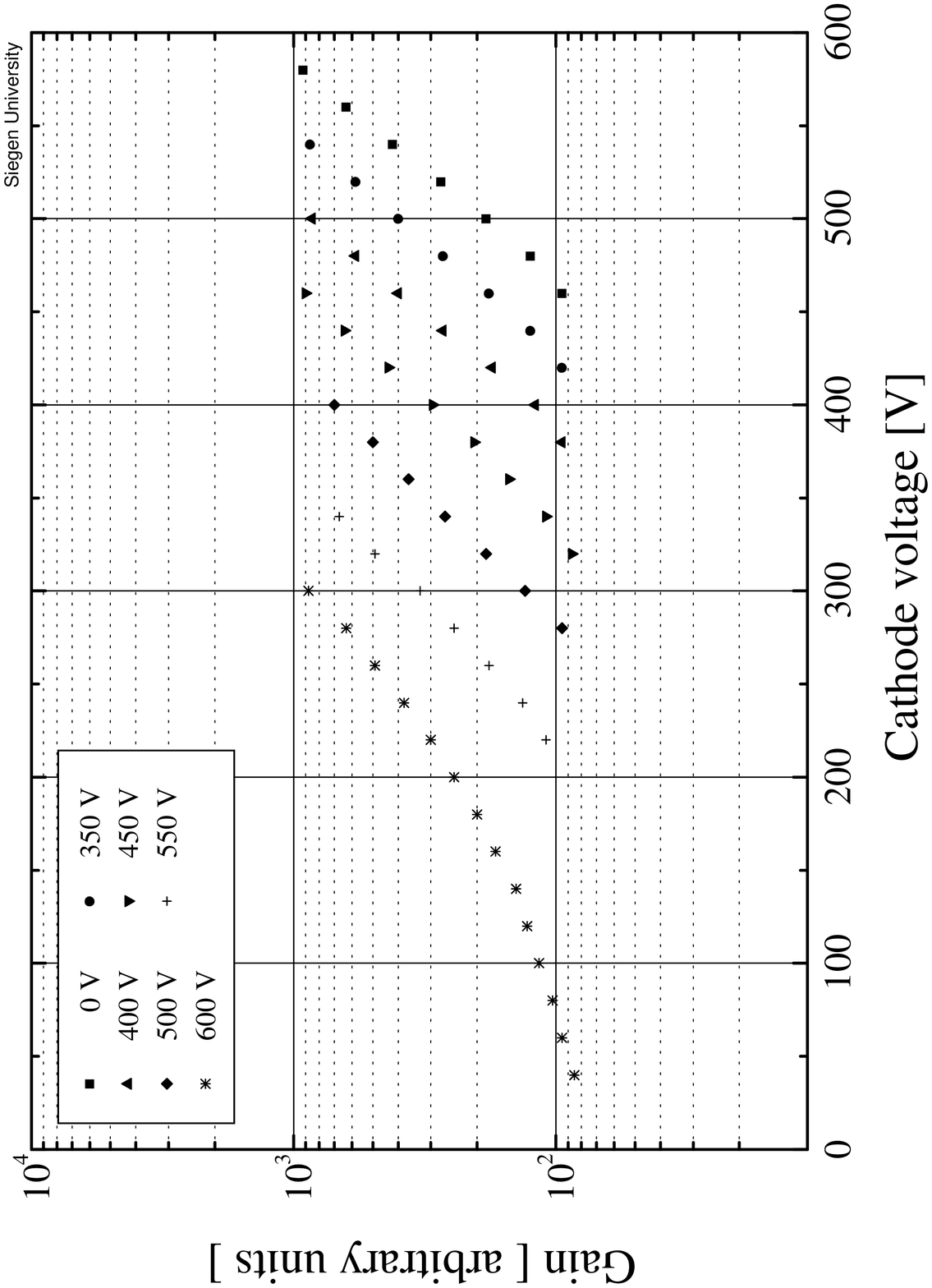}%
\caption{Gas amplification as a function of the MSGC cathode voltage for
different GEM voltages.}%
\label{gain_msgc_gem}%
\end{center}
\end{figure}
The leftmost curve is obtained with a cathode voltage of the MSGC of 50V only
which is needed solely to direct the electrons towards the anodes. The GEM can
safely be operated at gains of a few hundred. Both the coated and uncoated
GEMs start to produce discharges at amplification factors above 1000 (Figure
9).%
\begin{figure}
[ptbptb]
\begin{center}
\includegraphics[
height=17.9552cm,
width=12.7778cm
]%
{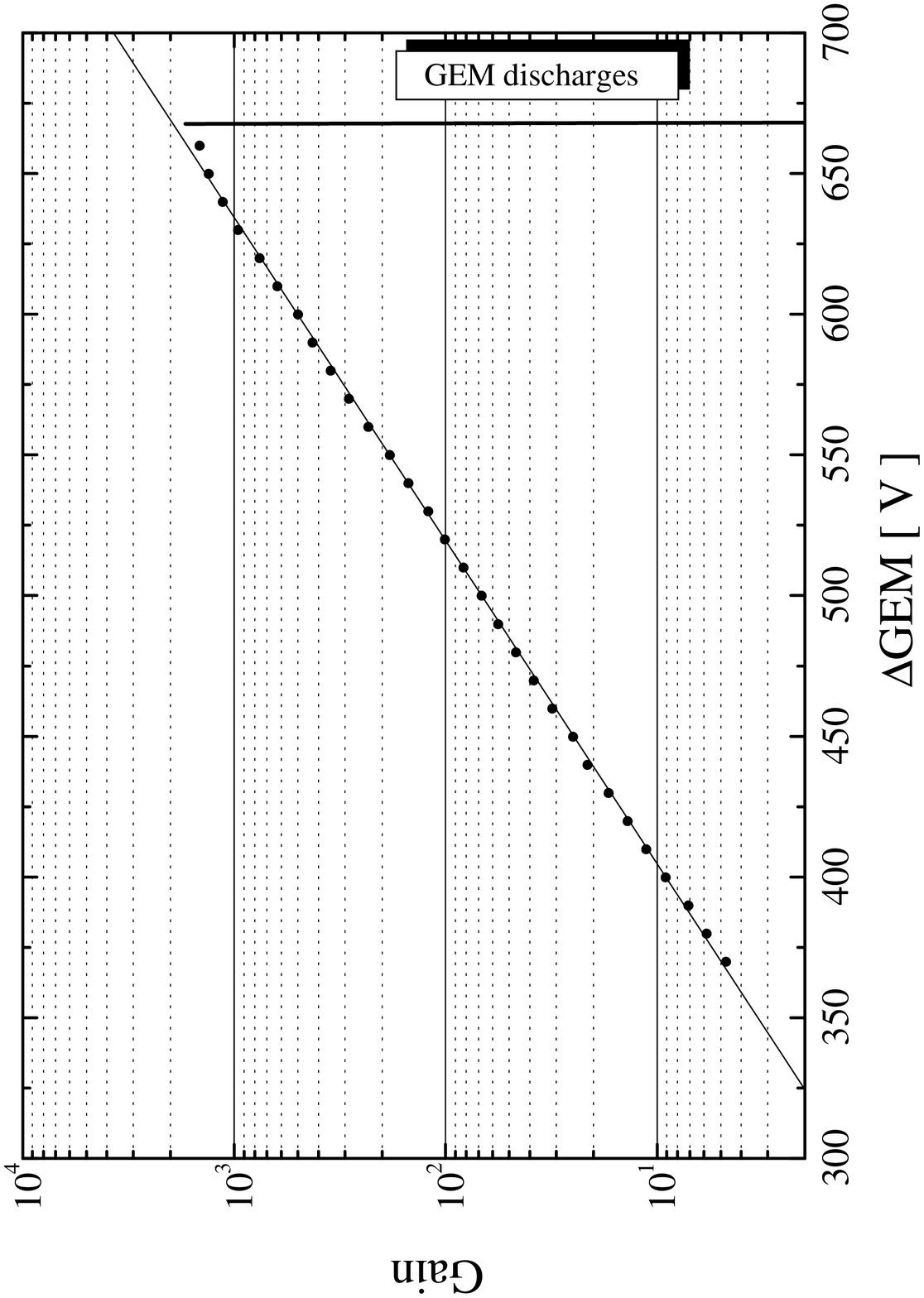}%
\caption{Gain of a coated GEM as a function of the applied voltage.}%
\label{gain_gem}%
\end{center}
\end{figure}
The discharges do not destroy or modify the performance of the GEM.

\subsection{Stability of the coating}

A coated GEM has been irradiated at a GEM voltage of $500V$ with an $\alpha$
source (Ra-226) of $5kHz$ per $10mm^{2}$ spot size for several days without
any deterioration, no discharges were observed. A long term aging test is
still required. However, experience with MSGCs indicate that coating should
rather improve the long term stability of gaseous detectors.

\section{Conclusions and outlook}

Coating of GEMs increases their performance. Instabilities due to charging up
and polarization are avoided. For a series of small GEMs with unusually bad
quality the coating improved the energy resolution and the spatial gain
uniformity. This observation may allow to relax the quality requirement for
the GEM production. The GEM quality of the larger GEMs (type L) we
investigated is much superior and their energy resolution and spatial gain
uniformity before coating was comparable to that of the coated GEMs.

At present we are studying GEMs with a $100\mu m$ thick polyimide layer. These
foils were produced at Moscow Lebedev Institute. Their first attempt to etch
the hole structure lead to a very bad quality. The copper sticks into the
holes thus the maximum achievable voltages are limited by discharges. However,
gains in excess of a few 1000 can be reached. We expect a new series with a
second etching step which should avoid discharges and allows to verify that
the coating increases the maximum achievable voltages and gains. We further
intend to apply the coating technique to a two-dimensional detector [10] where
the holes of the CAT [11] are replaced by grooves.

\textbf{Acknowledgement:} We are very grateful to Dr. F. Sauli for providing
us various GEM foils from the CERN workshop and for many critical comments to
our results.


\begin{thebibliography}{99}
\bibitem{1}F. Sauli, Nucl. Instr. and Meth. A386 (1997) 531

\bibitem{2}R. Bouclier et al., Nucl. Instr. and Meth. A 396 (1997) 50-66

\bibitem{3}B. Schmidt, HD-PY-98-02, Subm. 36th INFN Eloisatron Project
Workshop on New Detectors Nov. 1997

\bibitem{4}J. Benlloch et al., Vienna Wirechamber Conference 1998

\bibitem{5}F. Sauli, Vienna Wire Chamber Conference 1998

\bibitem{6}T. Zeuner, Nucl. Instr. and Meth. A392 (1997) 105

\bibitem{7}P. Fonte et al. submitted to Nucl. Instr. and Meth.

\bibitem{8}Y. Ivaniouchenkov et al., Symposium on Radiation Measurements and
Applications, Ann Arbor (1998)

\bibitem{9}S. Schmidt et al., Nucl. Instr. and Meth. A337 (1994) 386

\bibitem{10}S. Keller et al., Vienna Wire Chamber Conference 1998

\bibitem{11}F. Bartol et al., J. Phys. III, France 6 (1996) 337
\end{thebibliography}
\end{document}